\documentclass[pdflatex,sn-nature]{sn-jnl}

\usepackage{graphicx}%
\usepackage{multirow}%
\usepackage{amsmath,amssymb,amsfonts}%
\usepackage{amsthm}%
\usepackage{mathrsfs}%
\usepackage[title]{appendix}%
\usepackage{xcolor}%
\usepackage{textcomp}%
\usepackage{manyfoot}%
\usepackage{booktabs}%
\usepackage{algorithm}%
\usepackage{algorithmicx}%
\usepackage{textcomp,mathcomp}
\usepackage{algpseudocode}%
\usepackage{listings}%
\usepackage{array}
\usepackage{caption}

\theoremstyle{thmstyleone}%
\theoremstyle{thmstyletwo}%

\theoremstyle{thmstylethree}%

\begin{document}

\title[Article Title]{Novel active avalanche paradigm for power semiconductor devices extending SOA to 6.5kV/15kA}

\author[1,2]{\fnm{Liu} \sur{Jiapeng}}\email{liujiapeng@tsinghua.edu.cn}
\author[1,3]{\fnm{Liu} \sur{Fucheng}}\email{liufc24@mails.tsinghua.edu.cn}
\equalcont{These authors contributed equally to this work.}
\author*[1,3]{\fnm{Wu} \sur{Jinpeng}}\email{jinpengwu@tsinghua.edu.cn}
\author[1]{\fnm{Ren} \sur{Chunpin}}\email{rcp@tsinghua.edu.cn}
\author[1]{\fnm{Pan} \sur{Jianhong}}\email{pjh23@mails.tsinghua.edu.cn}
\author[1]{\fnm{Chen} \sur{Zhengyu}}\email{chenzhen14@tsinghua.org.cn}
\author[1,3]{\fnm{Zhao} \sur{Biao}}\email{zhao-biao@tsinghua.edu.cn}
\author[3]{\fnm{Li} \sur{Xiaozhao}}\email{lixiaozhao@neps.hrl.ac.cn}
\author[1,3]{\fnm{Zhuang} \sur{Chijie}}\email{chijie@tsinghua.edu.cn}
\author[1,3]{\fnm{Yu} \sur{Zhanqing}}\email{yzq@tsinghua.edu.cn}
\author[3]{\fnm{Wei} \sur{Xiaoguang}}\email{weixiaoguang@neps.hrl.ac.cn}
\author*[1,3]{\fnm{Zeng} \sur{Rong}}\email{zengrong@tsinghua.edu.cn}

\affil*[1]{\orgdiv{Department of Electrical Engineering}, \orgname{Tsinghua University}, \orgaddress{\city{Beijing}, \postcode{100084}, \country{China}}}

\affil[2]{\orgdiv{Department of Energy and Power Engineering}, \orgname{Tsinghua University}, \orgaddress{\city{Beijing}, \postcode{100084}, \country{China}}}

\affil[3]{\orgdiv{Next Generation Power System Research Center}, \orgname{Beijing Huairou Laboratory}, \orgaddress{\city{Beijing}, \postcode{100084}, \country{China}}}


\abstract{With the rapid growth of renewable energy being integrated, transmitted, and utilized in various forms, power systems are evolving from traditional metal-based infrastructures to advanced semiconductor-based technologies. Central to this transformation, power semiconductor devices now face the crucial demands of ultrahigh capacity, presenting a long-standing challenge in the field. In this work, building on the P-N-P-N scheme, we propose a novel Avalanche Controlled Thyristor (ACT) device, which pivots an active avalanche paradigm to realize ultrahigh capacity and robust operation safety simultaneously through semiconductor-level electrical potential manipulation. Unprecedentedly, we achieve a safe switch-off operation under load current of 15kA, 375\% that of the conventional devices without sacrificing the blocking and conduction capability. The novel device signifies that operation safety is no longer a bottleneck for power semiconductor devices, which is believed to revolutionarily reshape the technical and practical landscape of power conversion, and fundamentally boost the large-scale integration, cross-regional transmission, and cost-effective utilization of renewable energy.}

\keywords{Power semiconductor device, Energy system, Avalanche, Capacity, SOA}

\maketitle

\section{Introduction}\label{sec1}

\begin{figure}[p] 
    \centering
    \includegraphics[width=1\linewidth]{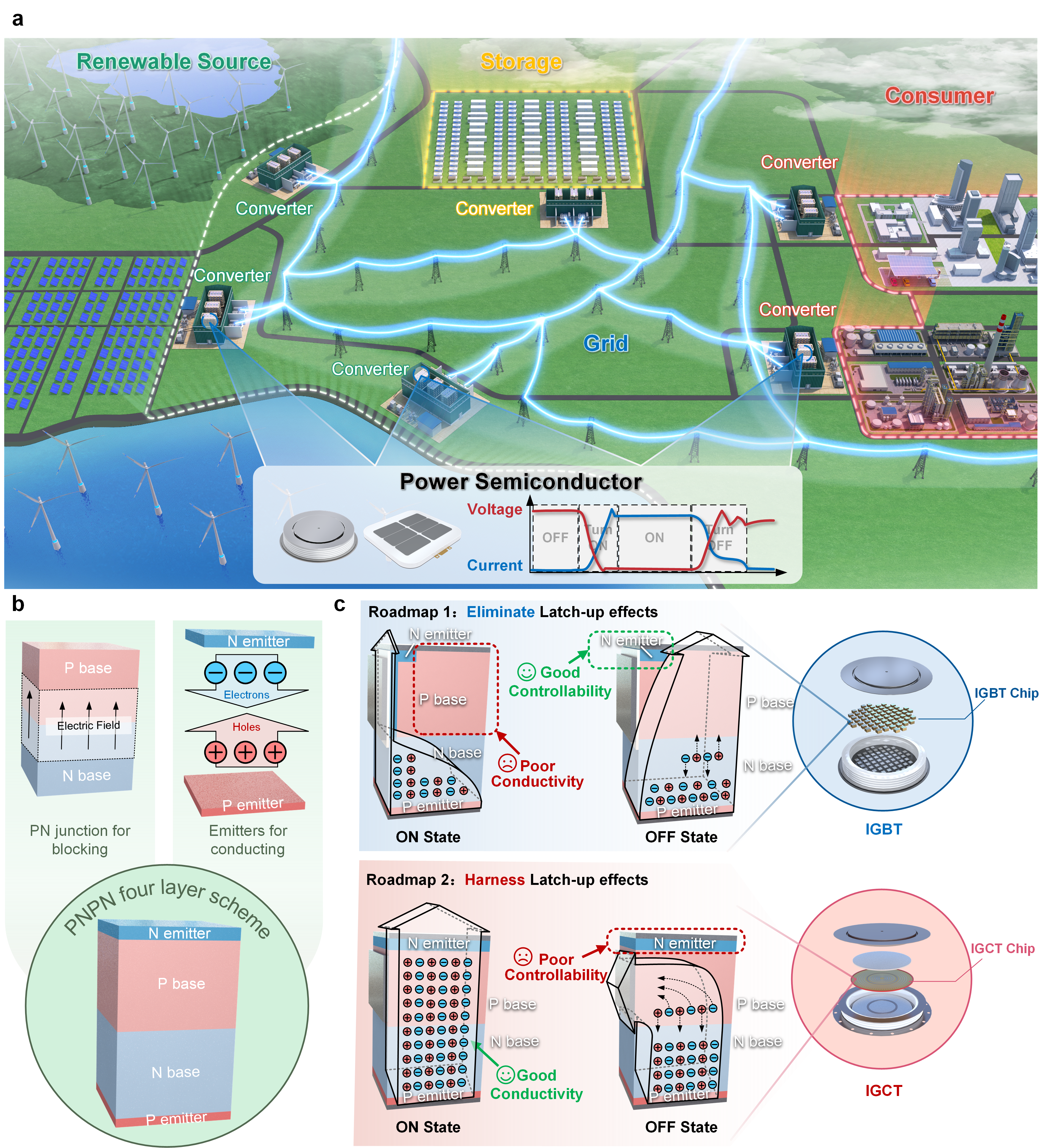}
    
    \caption{\textbf{Fig. 1 The Necessity and Challenges of High-Capacity Power semiconductors.}\\
\textbf{a,} The rapid development of renewable energy accelerates the wide deployment of semiconductor-based energy conversion equipment, for which the high capacity power semiconductor devices are of crucial significance.
\textbf{b,} The P-N-P-N four-layer structure has become key to meeting the requirement for both  high voltage blocking endurance and strong conduction capability in power semiconductor devices.
\textbf{c,} The challenge of mitigating latch-up modes has led to divergent development paths in modern high voltage power semiconductors: one path prioritizes safe operation by eliminating the latch-up mechanism, while the other embraces the latch-up mode to enhance capacity, albeit at the expense of operation safety.}
    \label{FIG1}
\end{figure}

In recent years, to encounter the challenges raised by climate change and air pollution, energy systems are evolving profoundly towards environmental friendliness\cite{wang2023accelerating}\cite{sovacool2020differences}\cite{guo2022implications}. With the rapid growth of renewable energy being integrated, transmitted, and utilized in various forms, a vast number of power semiconductor devices and energy converters are being deployed within the power systems\cite{zhang2022multidimensional}\cite{boroyevich2013intergrid}\cite{kroposki2017achieving}, which drives a swift upgrade of energy systems from metal-based to semiconductor-based technologies\cite{boroyevich2010future}\cite{zhao2018more}\cite{sajadi2022synchronization} (refer to Fig. \ref{FIG1}a).

In principle, power semiconductor devices serve the functions of blocking and conducting the power energy flow (namely OFF and ON states, respectively), and making a switch between the two states\cite{zhou2023avalanche}\cite{van2013future}. Thus, a well-performed power semiconductor device is generally characterized by high voltage blocking endurance and high conduction capability\cite{zhang2019novel} . Intrinsically, a dilemma lies in the OFF and ON states in terms of carrier density, where an insulator-like low density and a conductor-like high density are respectively preferred for the OFF and ON states\cite{ren2023optimal}. To resolve the dilemma, a lowly doped P-N junction (i.e. P and N bases) is firstly adopted for strong blocking during OFF state\cite{baliga2010fundamentals}; on this basis, highly doped hole and electron emitting regions (i.e. P and N emitters respectively) are then adopted for carrier injection during ON-state, forming strong conductivity modulation effect to enhance conduction\cite{liu2020ultra}. Thus, the four-layer P (emitter) - N (base) - P (base) - N (emitter) scheme is deemed as the optimal solution for achieving high blocking voltage and large capacity within one power semiconductor device\cite{huang2019power}\cite{zhang1997emitter} (refer to Fig. \ref{FIG1}b).

Under the P-N-P-N scheme, to obtain a supreme conduction performance and thus a sufficient device capacity, the latch-up mode plays a vital important role, for which the self-sustaining positive feedback between the two emitters guarantees a satisfying high carrier density for conduction\cite{mueller1958thyristor}. However, from another perspective, the extremely strong positive feedback also causes trouble in latch-up breaking and consequently switch-off failure\cite{baliga1996trends}. Addressing the conflict, one roadmap, represented by IGBT (invented in 1982\cite{baliga1982insulated}, refer to Fig. \ref{FIG1}c), chooses to circumvent the latch-up effect for the realization of operation safety\cite{temple1986mos}\cite{wang2022igbt}. As a consequence, the device capacity is restricted due to the lack of latch-up under the P-N-P-N scheme\cite{baliga1990mos}\cite{nandakumar1991new}.

In contrast, another roadmap, represented by IGCT (invented in 1997\cite{steimer1997igct}, refer to Fig. \ref{FIG1}c), harnesses the latch-up effect to capitalize its superior conduction performance, while leaving the operation safety issue not well resolved\cite{chen2017analysis}\cite{chen2024enhancing}. To be more specific, IGCT utilizes external circuit to tune the P-N junction between the cathode and gate negatively biased, aiming at ceasing carrier injection and breaking the latch-up reliably\cite{shang2022novel}. However, upon the load current increasing, the uneliminable parasitic impedance within the external circuit and semiconductor gradually depresses the stability of negative-bias state of P-N junction and ultimately flips the P-N junction back to positively biased status, leading to the carrier injection retrigger and the device destruction\cite{liu2024experimental}\cite{lophitis2013destruction}. Though many optimization efforts have been made to suppress the parasitic parameters\cite{chen2018stray}\cite{zhou2021p+}\cite{shang2023novel}, this intrinsic issue characterized by insufficient commutation capability will inevitably occur within IGCT.

In this work, building on the P-N-P-N scheme, we propose a novel Avalanche Controlled Thyristor (ACT) device, which not only leverages the latch-up effect to achieve superior capacity, but more importantly, pivots an active avalanche paradigm to realize robust operation safety. Through discerning manipulation of electrical potential inside semiconductor, the paradigm is capable of diminishing the impact of non-ideal external parasitic factor. As a consequence, the paradigm thoroughly unleashes the safe operation area for latch-up devices, and succeeds in realizing over a 275\% improvement in the safe operation current without sacrificing conduction performance in comparison with the conventional devices, marking the unprecedented advancement in harmonizing IGBT-like operation safety and IGCT-like capacity in the power semiconductor field. We believe that the invention of the ACT device will ultimately overcome the bottleneck in the operation safety and capacity of power semiconductor devices, fundamentally revolutionizing the technical and practical landscape of power conversion, and finally boosting the large-scale integration, cross-regional transmission, and economical utilization of renewable energy.

\section{Paradigm Upgrade: from Reverse Bias to Controlled Avalanche}\label{sec2}

\begin{figure}[p] 
    \centering
    \includegraphics[width=1\linewidth]{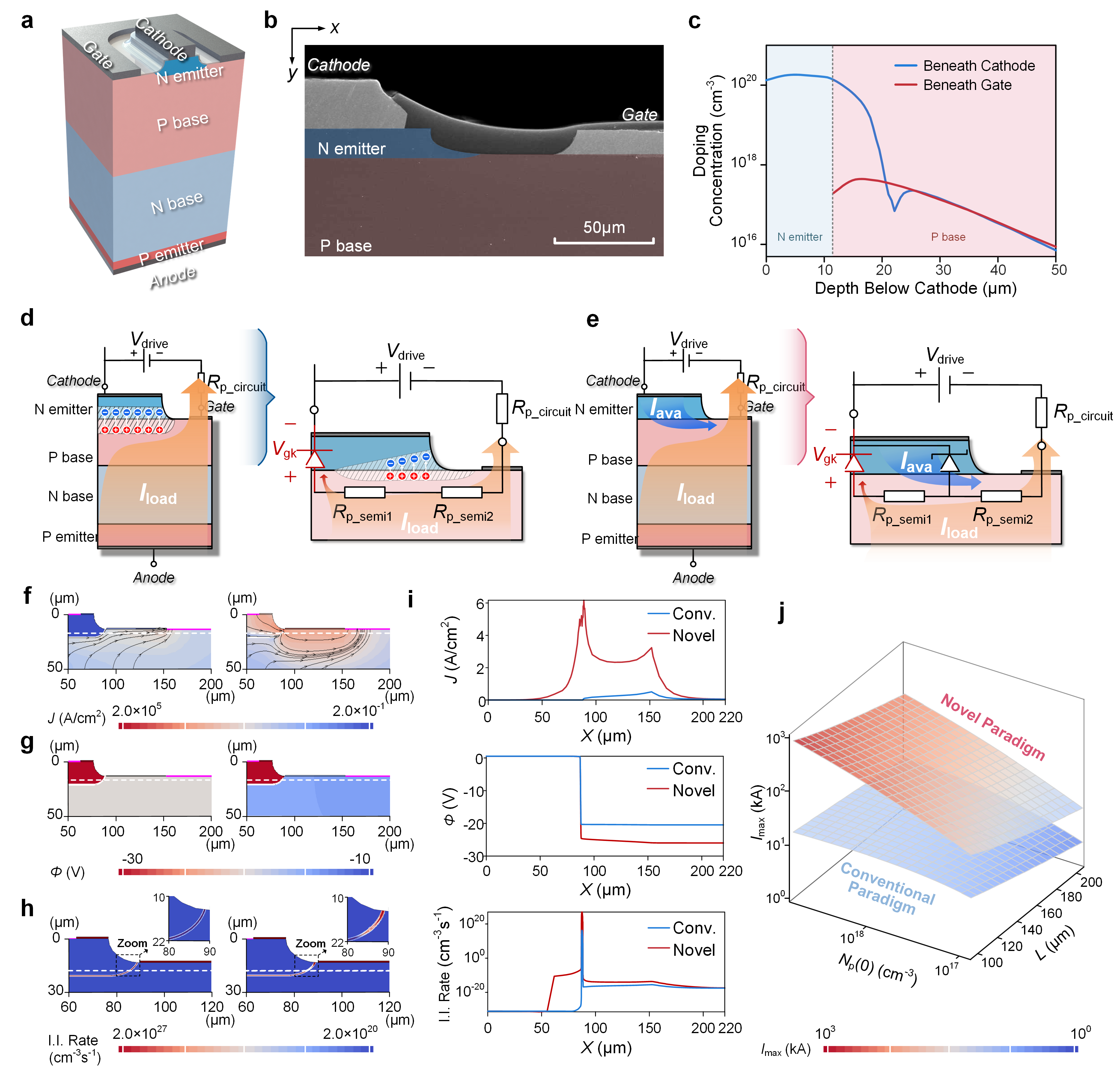}
    \caption{\textbf{Fig. 2 The demonstration of fundamental principle and critical characteristics for controlled avalanche paradigm.}\\
\textbf{a,} Schematic of a fundamental cell for latch-up devices.
\textbf{b,} Cross-sectional SEM image of latch-up devices.
\textbf{c,} Doping profile of the emitter junction.
\textbf{d,} Operation principle (left) and critical failure state (right) for conventional paradigm (reverse bias).
\textbf{e,} Operation principle (left) and critical failure state (right) for novel paradigm (controlled avalanche).
\textbf{f,} Simulated current density  ($J$) distribution during switching off under conventional and novel paradigm (black lines represents the current streamlines; Left: reverse bias paradigm; Right: controlled avalanche paradigm ).
\textbf{g,} Simulated  electrostatic potential ($\phi$) distribution during switching off under conventional and novel paradigm (Left: reverse bias paradigm; Right: controlled avalanche paradigm).
\textbf{h,} Simulated  impact ionization rate (I.I. Rate) distribution during switching off under conventional and novel paradigm (Left: reverse bias paradigm; Right: controlled avalanche paradigm).
\textbf{i,} Simulated profiles of the current density, electrostatic potential and impact ionization rate at \textit{Y}=18$\mathrm{\mu m}$
\textbf{j,} The theoretical maximum safe operation current achievable through the controlled avalanche paradigm.}
    \label{FIG2}
\end{figure}

The structure and doping profiles of a fundamental cell within the latch-up type P–N–P–N semiconductor is shown as Fig. \ref{FIG2}a,b,c. As aforementioned, the utmost challenge to switch off a latch-up device is to cut off the carrier injection of the emitter junction concretely. To achieve this, the conventional paradigm adopts the strategy of compelling the emitter junction reverse bias. After activating the voltage source $V_{drive}$ (lower than junction avalanche voltage $V_{ava}$) in gate driver, as demonstrated in the left of Fig\ref{FIG2}d, the load current commutate and redirect to gate and the depletion region is established in the emitter junction, thus the carrier injection from N emitter can be eliminated. Under this paradigm, taking account the parasitic resistances both in the driver circuit and inside the semiconductor, the voltage genuinely applied at the emitter junction $V_{gk}$ can be denoted as:
\begin{equation}
    V_{gk} =-V_{drive}+\underbrace{I_{load}R_{p\_circuit}}_{V_{p\_circuit}}+\underbrace{I_{load}(R_{p\_semi1}+R_{p\_semi2})}_{V_{p\_semi}}.
    \label{eq1}
\end{equation}
where $R_{p\_circuit}$ represents for the parasitic resistance in the drive circuit and can be regarded as a macroscopic constant, $R_{p\_semi1}$ and $R_{p\_semi2}$ represent for the microscopic equivalent lateral resistances in the P base region beneath the N emitter and the spacer, respectively (here for simplification, the inductance effect is omitted; refer to Supplementary Sec. S1 for the detailed deviation).

Based on Eq.(\ref{eq1}), it can be deduced that with $I_{load}$ increasing, the junction bias voltage $V_{gk}$ keeps decreasing. Until $I_{load}$ reaches up to the critical value of $I_{max}={V_{drive}}/({R_{p\_circuit}+R_{p\_semi1}+R_{p\_semi2}})$, the emitter junction will be re-diverted from reverse bias into forward bias, leading to uncontrolled re-latch-up (as demonstrated by the right of Fig. \ref{FIG2}d), indicating the occurrence of switch-off failure and consequent destruction. 

Fig. \ref{FIG2}j provides an visualized estimation on $I_{max}$ for a 4-in-diameter device with two most crucial semiconductor-level design parameters $L$ , defined as the linewidth of emitter junction, and surface doping concentration $N_P(0)$ as variables (refer to Methods for detailed parameter configurations). As can be concluded, though fine tuning of parameters may help improving $I_{max}$ (usually at the cost of process complexity or conduction performance), the existence of unavoidable parasitic resistances in the external circuit and inside the semiconductor still hinders further improvement of $I_{max}$ for the conventional paradigm.

To acquire a leap enhancement in switch-off operation safety, we need to jump outside the box. Under conventional paradigm, $V_{drive}$ is severely restricted not too high to avoid the occurrence of avalanche on the emitter junction, as the avalanche is conventionally deemed detrimental in terms of reliability and stability. With such limitation on $V_{drive}$, the reverse bias status is substantially fragile upon large load current, due to the influence of the parasitic resistances. Over the past several decades, both academia and industry have well accepted this paradigm\cite{zhou2021p+}\cite{shang2023novel}\cite{steimer2001igct}\cite{gruening2005modern}. 

However, once reverse avalanche is permitted by purpose, the constraint on $V_{drive}$ will be unlocked, and can be exploited to a relatively high value to assure the reverse avalanche status on the emitter junction, as demonstrated in the left of Fig. \ref{FIG2}e. Under these conditions, the emitter junction transitions into a fundamentally different operational state, entering a reverse avalanche regime. At this point, $V_{gk}$ is pinned steadily to the constant of $-V_{ava}$, irrelevant of load current. This suggests that the detrimental voltage drops induced by load current across parasitic resistances can be almost mitigated, enabling a substantial improvement in the switch-off safety of latch-up devices. 

Quantitatively, simulations were performed to compare the actual operating state of the emitter junction under conventional and novel paradigms, focusing on current density, electrostatic potential and impact ionization rate. The results, depicted in Fig. \ref{FIG2}f-i, reveal that under the novel paradigm, the periphery of the emitter junction enters a reverse avalanche state, where a high local impact ionization rate is sustained and the junction voltage bias is effectively pinned. This localized avalanche behavior plays a critical role in stabilizing the junction potential during switch-off process. Consequently, in evaluating the maximum switch-off current under novel paradigm, the influence of the external circuit has been disregarded, while the parasitic resistance inside the semiconductor still needs to be taken into consideration. Under critical switch-off load conditions, though the periphery position of the emitter junction still fixes at the avalanche status, the central position of the emitter junction may recover from the avalanche status and turns back to forward bias status, leading to a re-latch-up failure of the novel paradigm, as demonstrated by the right of Fig. \ref{FIG2}e .  $V_{gk}$ and $I_{max}$ can thus be quantitatively evaluated as following:
\begin{equation}
    V_{gk} =-V_{ava}+\underbrace{I_{load}R_{p\_semi1}}_{V_{p\_semi1}}.
    \label{eq2}
\end{equation}
\begin{equation}
   I_{max}=\frac{V_{ava}}{R_{p\_semi1}}.
    \label{eq3}
\end{equation}

For an intuitive comparison, the critical load currents for both paradigms are depicted in Fig. \ref{FIG2}j. As can be told, approximately and theoretically, more than 400\% improvement can be obtained with proposed paradigm, marking a revolutionary breakthrough in the aspect of device operation safety.

\section{Avalanche Stability Analysis and Stability Boundary}\label{sec3}

For the novel paradigm, one of the main concerns for its feasibility is raised by stability of junction avalanche under load-level avalanche current, which is obviously not a common operation mode for high power devices. Under such circumstance, the overheating phenomenon, especially within a small portion of the device, may occur and destroy the semiconductor permanently. To address this issue, the balanced current sharing and consequently temperature equalizing should be guaranteed firstly as stability precondition. Based on that, the safe operation boundary could be depicted in prevention for monolithic overheating. 

\begin{figure}[p] 
    \centering
    \includegraphics[width=1\linewidth]{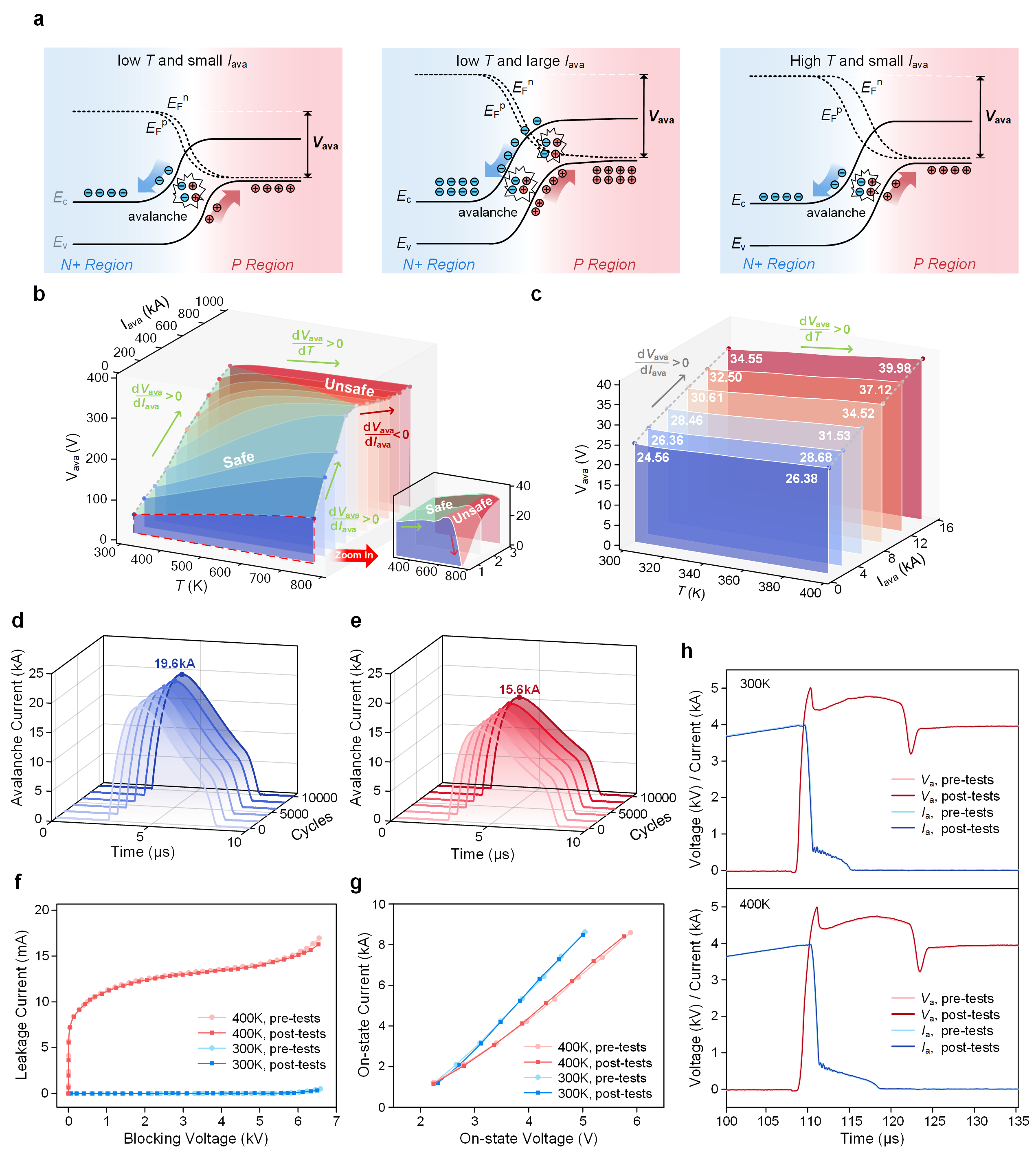}
    \caption{\textbf{Fig. 3 Stability assessment and reliability validation of the avalanche mechanism.}\\
\textbf{a,} The evolution of band structure and carrier dynamics with varying current and temperature.
\textbf{b,} TCAD simulations show the avalanche remains stable up to 650K temperature and 400kA avalanche current.
\textbf{c,} Experimental results confirm junction stability at temperatures up to 400K and avalanche currents up to 15kA.
\textbf{d,} Sampled waveform of avalanche current from 10,000 test cycles at 300K.
\textbf{e,} Sampled waveform of avalanche current from 10,000 test cycles at 400K.
\textbf{f,} Comparison of blocking characteristics pre- and post- avalanche cycling test.
\textbf{g,} Comparison of conduction characteristics pre- and post- avalanche cycling test.
\textbf{h,} Comparison of dynamic characteristics pre- and post- avalanche cycling test.}
    \label{FIG3}
\end{figure}
In terms of current balance, the feature is based predominantly on the negative feedback mechanism founded on the positive correlation of avalanche current $I_{ava}$ and temperature $T$ with avalanche voltage $V_{ava}$. 
\begin{equation}
   \frac{\partial V_{ava}}{\partial I_{ava}}>0 .
    \label{eq4}
\end{equation}
\begin{equation}
   \frac{\partial V_{ava}}{\partial T}>0.
    \label{eq5}
\end{equation}

Based on this, the region experiencing locally elevated current or temperature develops a higher avalanche voltage, thereby suppressing transient accidental fluctuation in turn. Fortunately, this behavior is rooted in the self-stabilizing nature of avalanche processes, as depicted in Fig. \ref{FIG3}a. Under a high avalanche current $I_{ava}$ or a higher temperature $T$, the maximum electric field and thus the avalanche voltage $V_{ava}$ increase\cite{okuto1975threshold}, until the generated mobile charges approach the doping level and severe distortion occurs of the electric field (see Supplementary Sect. S2 for a detailed explanation). Therefore, Eq.(\ref{eq4}) and (\ref{eq5}) naturally hold within a wide operating range. 

In order to provide better demonstration to the stability precondition for a typical silicon P-N-P-N structure under conventional paradigm, the impact of $I_{ava}$ and $T$ on $V_{ava}$ is depicted on Fig. \ref{FIG3}b based on simulation. As can be seen, the junction avalanche is stable until $T$ reaches $T_{crit}$ = 650K or $I_{ava}$ reaches $I_{ava\_crit}$ = 400kA.

Another factor that needs to be considered is the thermal restriction for monolithic overheating, which is basically decided by temperature $T_{d}$, which may cause permanent dopant re-diffusion. Specifically, for silicon, $T_{d}$ can be considered higher than 800K\cite{krause2002determination}, which is still much higher than $T_{crit}$, indicating that for the emitter junction, the theoretical safe operation region is fully determined by the stability precondition. 

Concerning that the avalanche paradigm is only activated at switch-off process, the boundary in the form of the driving voltage $V_{drive}$ and the avalanche period $t_{ava}$ can be obtained by simulation (refer to Supplementary Sec. S3 for details). The parameters are chosen as fixed values of 56V and 4.4µs in this work to overlap the whole switch-off dynamics while providing sufficient safe margin.

To verify the theoretical deduction on avalanche stability and further the feasibility of proposed paradigm in real applications, both the measurement on $V_{ava}-I_{ava}$ relation and short-term reliability test on controlled avalanche are performed. Fig. \ref{FIG3}c demonstrates the measured $V_{ava}-I_{ava}$ relation under various operation temperatures (refer to Methods for detailed configurations). As can be told, the values of ${\partial V_{ava}}/{\partial I_{ava}}$ and ${\partial V_{ava}}/{\partial T}$ are positive as expected, proving the self-stabilizing mechanism in emitter junction and consequently paradigm stability within $I_{ava} \leq$15kA and $T\leq$400K. Based on that, the short-term reliability tests with zero load current are performed for 10,000 times to emulate the most rigorous application condition with maximum avalanche current, as demonstrated by Fig. \ref{FIG3}d,e. The main electrical characteristics are measured pre- and post- the avalanche pulses to evaluate the existence of aging problem, as illustrated in Fig. \ref{FIG3}f,g,h. As can be told, no observable change or parameters shift can be observed for the test, primarily validating the feasibility of proposed paradigm.

\section{Characterization of ACT}\label{sec4}

\begin{figure}[t!] 
    \centering
   \includegraphics[width=1\linewidth]{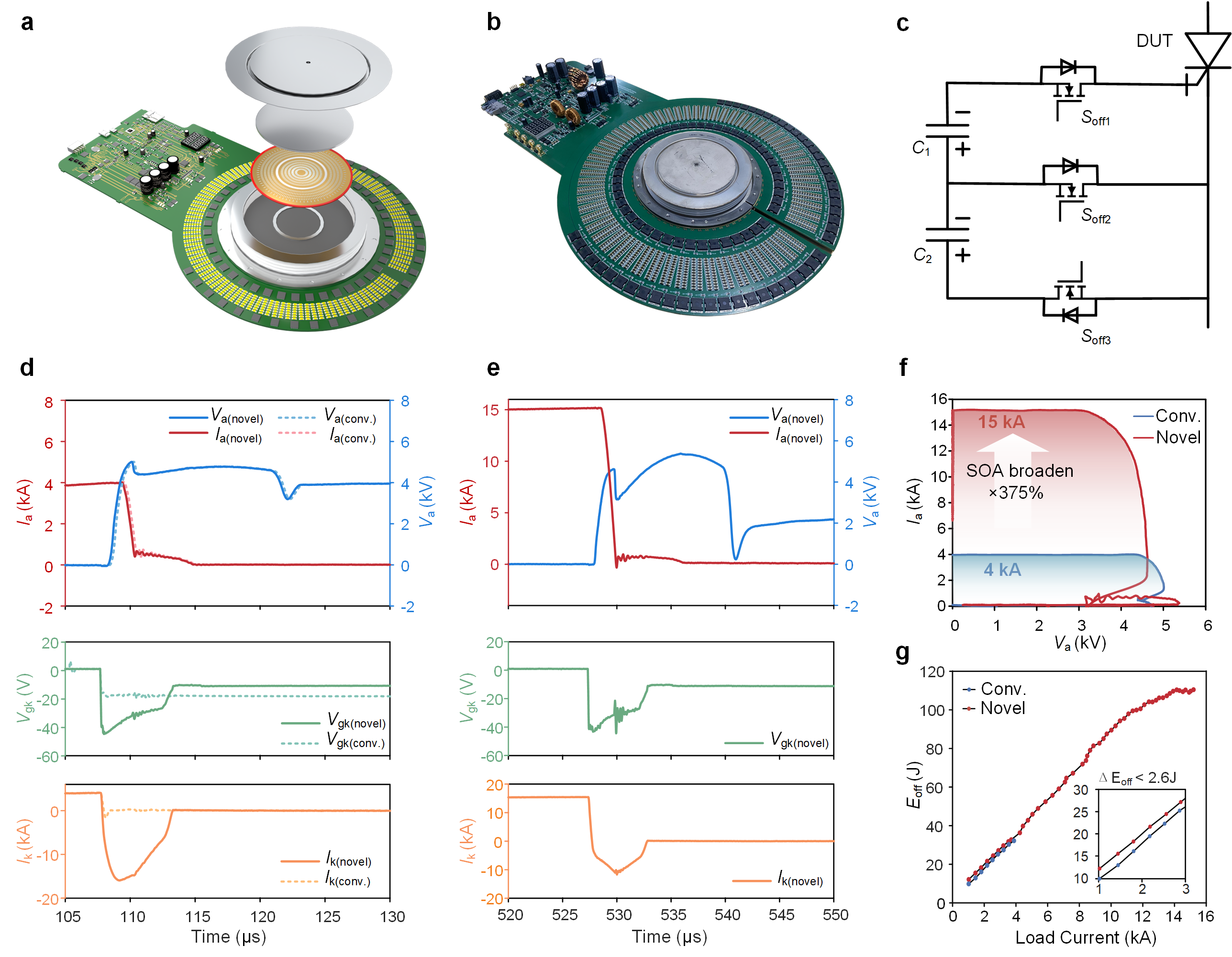}
    \caption{\textbf{Fig. 4 Characterization of the ACT.}\\
\textbf{a,} Schematic diagram of the ACT device.
\textbf{b,} Prototype of ACT device.
\textbf{c,} Topology of two-level gate drive circuit to implement controlled-avalanche.
\textbf{d,} Comparison of the switch-off waveforms of novel ACT devices and conventional latch-up devices (IGCT).
\textbf{e,} Switch-off waveforms of the ACT under 15kA load current.
\textbf{f,}Comparison of the safe operating area (SOA) between the novel paradigm and the conventional approach.
\textbf{g,} Switch-off energy loss comparison between the novel and conventional paradigm.}
    \label{FIG4}
\end{figure}

Based on the proposed paradigm, a novel latch-up P-N-P-N device with robust operation safety is presented, termed as Avalanche Controlled Thyristor (ACT). Its structural schematic is shown in Fig. \ref{FIG4}a.  To certificate the advancement of novel device, an ACT prototype with 4-in-diameter chip is fabricated, as demonstrated by Fig. \ref{FIG4}b. The avalanche-controlled switching mechanism is realized via a two-level gate drive circuit (Fig. \ref{FIG4}c), designed to meet the condition $V_{C1}<V_{ava}$ and $V_{C1}+V_{C2}>V_{ava}$,  thereby enabling controlled activation of the avalanche process only during switch-off. 

For demonstration conciseness, the characterization in this section focuses more on differentiated features in comparison with conventional latch-up devices (i.e. IGCT), which is mainly switch-off characteristic. To emulate the real power conversion situation, a typical phase-leg circuit is adopted for single pulse test (refer to Methods for detailed configurations). In order to verify the successful implementation of the paradigm, experiments were carried out within the rated safe operating range. As demonstrated  by Fig. \ref{FIG4}, the prototype adopting novel controlled avalanche paradigm exhibits external characteristics comparable to those of conventional approaches. 

For the aspect of safe operation area, as demonstrated by Fig. \ref{FIG4}e, the ACT sample finally succeeds in switching off a load current of 15kA without failure, which is approximately 4 times the rating capability of IGCT under conventional paradigm. No further test is performed just because anode voltage reaching close to zero due to circuit oscillation. As can be told from the waveform, the reverse current flows into cathode during switch-off dynamics, indicating a successful implementation of novel paradigm. If we convert the waveform from time domain into I-V domain, as depicted in Fig. \ref{FIG4}f, a quite explicit extension in safe operation area could be observed.

Fig. \ref{FIG4}g presents the measured switch-off energy loss of ACT, which may be of much concern owing to the existence of junction avalanche and consequently extra losses. As can be seen, the measured additional loss is less than 2.6J under various application scenarios, constituting less than 8\% of the switch-off losses (under rating current). The marginal increase can be regarded acceptable for ultra-high power applications.

\section{Discussion}\label{sec5}
\begin{figure}[t!] 
    \centering
    \includegraphics[width=1\linewidth]{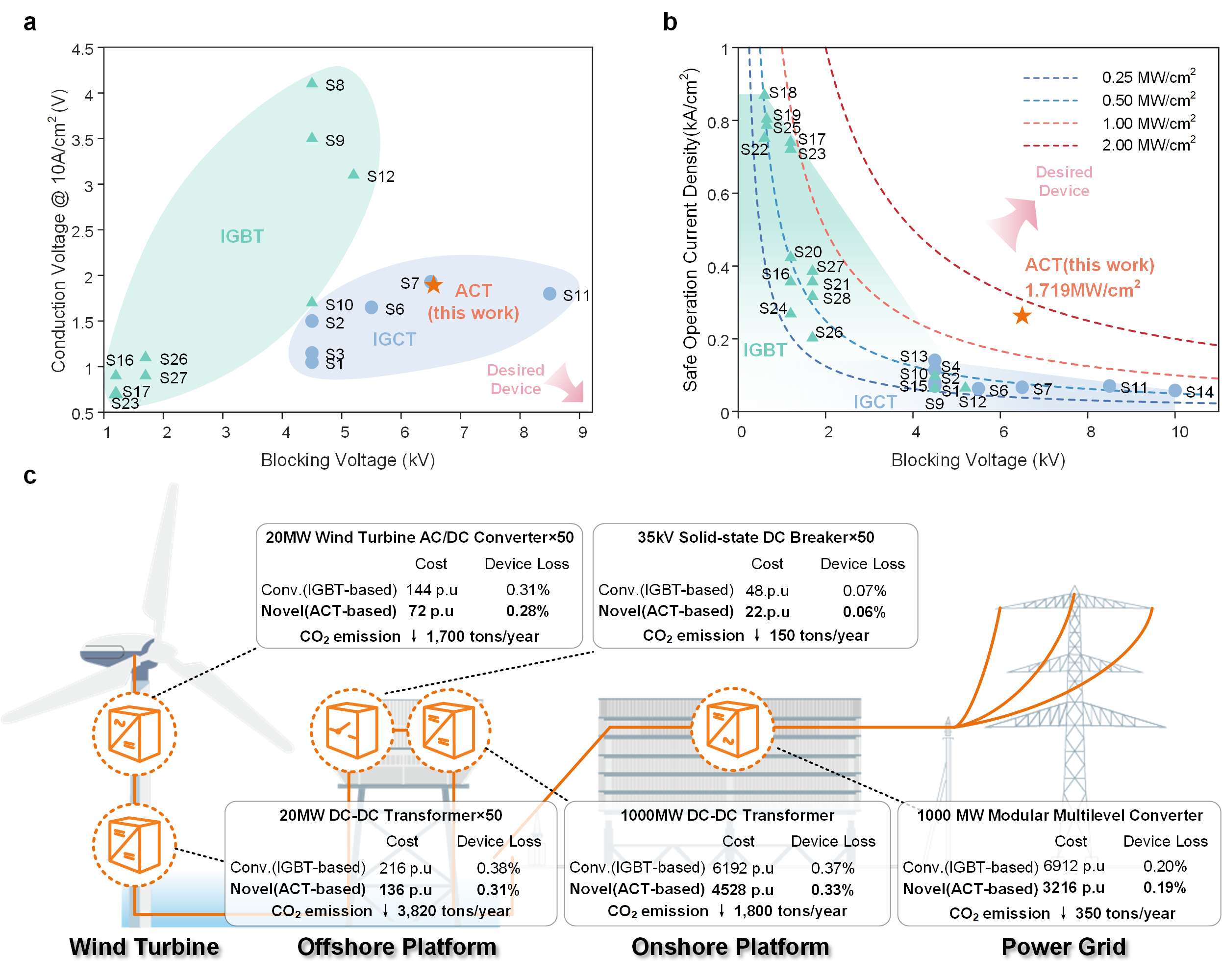}
    \caption{\textbf{Fig. 5 Capacity–SOA landscape and practical demonstration of ACT in a representative application case .  }\\
\textbf{a,} Comparison of ACT, IGBT and IGCT on capacity performance, demonstrating that ACT possesses the capacity advantage of latch-up devices.
\textbf{b,} Comparison of ACT, IGBT and IGCT on SOA performance, demonstrating that ACT extents greatly the SOA of latch-up devices.
\textbf{c,} Comparative analysis of 1000MW offshore wind farms employing various device techniques 
}
    \label{FIG5}
\end{figure}
To demonstrate the capacity and SOA advantages of ACT, we compare its performance with reported silicon-based high voltage semiconductor devices in Fig. \ref{FIG5}a,b . Our work, just implemented on the blocking voltage rating of 6.5kV, stands out as the only silicon-based device capable of handling power densities exceeding 1MW/cm$^2$ and close to 2MW/cm$^2$ without reaching its true limit, resolving the longstanding challenge harmonizing device capacity and SOA for ultra-high power semiconductor devices. It is worth noting that the ACT presented in this work is merely a preliminary prototype and is expected to ignite new research and development trends in the power semiconductor field. These trends may include, but are not limited to, further optimization of the ACT device structure and expanded use of wide band-gap materials such as Silicon Carbide, Gallium Oxide, diamond, and others.

The advent of ACT technology brings up several exciting new possibilities for future. The leapfrogging advancement of the performance significantly enhances the capacity and efficiency of wind turbine converters\cite{blaabjerg2013future}\cite{blaabjerg2024power}, photovoltaic inverters\cite{araujo2009highly}, AC/DC interconverters\cite{huber2017applicability}\cite{she2013review}\cite{nami2014modular}, etc. enabling the widespread economic utilization of abundant renewable energy, pushing forward the fully realization of renewable energy power supply, as demonstrated by a typical 1000 MW offshore wind farm in Fig.\ref{FIG5}c. 

By 2030, the total capacity of renewable energy is supposed to reach 10 TW\cite{iea2024renewables}. In the collection and transmission of renewable energy, the adoption of ACT technique is estimated to provide over 800 billion dollars saving in construction cost, and over 20 million tons of $CO_2$ emission reduction annually. This dual advantage underscores ACT not merely as a technological breakthrough, but as a pivotal enabler in accelerating the deployment of renewable energy and advancing the global agenda toward carbon neutrality.

Furthermore, the exceptional performance of the power semiconductor device decreases the technical difficulties and manufacturing costs of semiconductor-based equipment, practically accelerating the realization of a completely new form of high-resilient high-flexibility power energy system\cite{hannan2020state}\cite{kassakian2013evolving}. Besides, the introduction of ACT device may also advance aerospace industry \cite{doyle1995electromagnetic}\cite{fair2009advances} and fusion energy utilization\cite{yao2012advanced}, making ideas of high power electromagnetic launch system and controllable nuclear fusion possible reality in the near future.

\bibliography{sn-bibliography}
\clearpage
\makeatletter
\renewcommand\@biblabel[1]{[50]}
\makeatother
\section{Methods}\label{sec6}

\subsection{Semiconductor Chip Preparation}\label{subsec1}

The P-N-P-N semiconductor chip with blocking capability of 6.5kV in this work is developed with a 720µm, high-resistivity ($\sim$500$\Omega$·cm ) silicon wafer. The doping profile is measured with SSM 2100. Considering the diffusion can be approximately regarded as constant total dopant diffusion, the doping concentration $N(y)$ can be described with Gaussian distribution $N(y)=N_s\exp{(-(y-y_0)^2/2L_0^2)}$, where $N_s$, $y_0$  and $L_0$ represent for peak doping concentration, symmetry depth and normalized depth respectively. Extended Data Table 1 demonstrates the doping profiles in different regions formed by diffusion.

\subsection{Theoretical Limit Evaluation of $I_{max}$ under conventional and novel paradigm}\label{subsec2}

For conventional paradigm, the evaluation of $I_{max}$ is based on (\ref{eq1}) with the expression of $R_{semi1}+R_{semi2}$ as following (refer to Supplementary Sec. S1 for deviation):
\begin{equation}
R_{p\_semi1}+R_{p\_semi2}
    = \frac{M(L)}{q S_{device}\int_0^D N_P(y)\mu_P(y)dy}.
    \label{eq6}
\end{equation}
Thus, the parameters of $M(L)$, $S_{device}$, $N_P$, $\mu_P$, $R_{circuit}$ and $V_{drive}$ need to be carefully selected for credible estimation.

The current density $J(x)$ is considered as the worst case for switch-off safety, where all current is crowded into a small portion at the center of cell and passing laterally through whole p base region to reach gate electrode. Under this circumstance, $J(x)$ is supposed to be defined in form of Dirac delta function as  $J(x)={I_{max}L_{cell}}\delta (x)/{S_{device}}$. Consequently, $M(L)$ is calculated as $L_{cell}L$ ($L_{cell}$ changes simultaneously with $L$ as variable with constant ratio of 1.2).

The device area $S_{device}$ is selected for 4-in-diameter wafer as 40.5cm$^2$, and doping concentration distribution in P base region $N_P(y)$ is described as $N_P(y)=N_P(0)\exp{(-y^2/2 {L_P} ^2)}$ with $L_P$=104.90µm, which is well consistent with typical dopant distribution measured with SSM 2100, as demonstrated in Extended Data Table 1 and Extended Data Fig. 1a. $\mu_P(y)$ is calculated with $N_P(y)$ based on Arora's model\cite{arora1982electron}. The parasitic resistance $R_{p\_circuit}$ and driving voltage $V_{drive}$ are selected based on typical measured value of 1.25m$\Omega$ and 20V respectively.

For novel paradigm, the evaluation of $I_{max}$ is based on Eq.(\ref{eq3}) with  the expression of $R_{semi1}$ as following (refer to Supplementary Sec. S1 for deviation):
\begin{equation}
R_{p\_semi1}
    = \frac{M(L_N)}{q S_{device}\int_0^D N_P(y)\mu_P(y)dy}.
    \label{eq7}
\end{equation}
In addition to the parameters stated above, the lateral width of N emitter $L_N$ is estimated as 0.55$L$.  . The avalanche voltage $V_{ava}$ is evaluated as 23V according to the measured I-V curve as demonstrated in Fig. \ref{FIG3}b.

\subsection{Realization of Controlled Avalanche}\label{subsec3}

Extended Data Fig. 1a illustrate the control algorithm of gate driver circuit realizing controlled avalanche. During switching off, both $S_{off1}$ and $S_{off3}$ are controlled ON, thus the driving voltage $V_{drive}=V_{C1}+V_{C2}$ higher than avalanche voltage $V_{ava}$, indicating the activation of avalanche. After the switch-off dynamics, $S_{off3}$ is switched OFF and $S_{off2}$ is switched ON. At this time, the driving voltage $V_{drive}=V_{C1}$ is decreased lower than $V_{ava}$, thus a reverse bias voltage is applied to ensure blocking status while avalanche is de-activated to avoid thermal instability. The duration of avalanche mode is fixed as 4.4µs in this work, which is sufficient to cover whole switch-off dynamics.

In hardware design, the components $C_1$ and $C_2$ are designed with capacitance of 66.55mF and 10.62mF respectively and are realized with ceramic capacitors to minimize parasitic impedance. The active switch $S_{off1}$, $S_{off2}$ and $S_{off3}$ are all composed of 49 N-channel MOSFET (MOT7146T) connected in parallel to satisfy capacity requirement and reduce parasitic resistance. For the convenience of voltage adjustment, the charging circuit is achieved with buck and boost converter with digitally controlled potentiometer, enabling real-time adjustment through optical fiber connected to host computer under customized communication protocol.

\subsection{Experiment Setup}\label{subsec4}

For the ease of comparison among devices, the test circuit adopts a standard hard-switch phase-leg topology as demonstrated in Extended Data Fig. 1b. For the data acquisition, the anode voltage and gate voltage are measured with Pintech PT-5230 high voltage differential probe and Tektronix TPP 1000 probe respectively. The anode current and gate current are measured with PEM CWT 150B and Cybertek CP9121S Rogowski coil respectively. The data is recorded by Tektronix MSO58B oscilloscope under the sample rate of 125M/s. The layout of experiment platform are illustrated by Fig. Extended Data Fig. 1c.
\renewcommand{\refname}{Method References}
\

\section{Data availability}
Source data supporting the findings of this study are provided with this paper.

\section{Acknowledgments}
This work is supported by the National Natural Science Foundation of China (52177153, 52207164, U23B20136) and Shui Mu Tsinghua Scholar Project.

\section{Author contributions}
J.L., F.L., J.W., C.R. conceived the project; J.L., F.L., C.R. and X.L. developed the device; F.L., Z.C., B.Z., Z.Y. and X.W. designed and performed the verification experiments; J.L, J.P. and C.Z. did the theoretical derivation and simulation; F.L. and J.W. wrote the paper with inputs from all the authors. The work was supervised by J.W. and R.Z..

\section{Competing Interests} 
The authors declare no competing interests.

\section{Additional Information} 
Supplementary Information is available for this paper.

\clearpage
\begin{table}[!t]
\raggedright
    \caption{\textbf{Extended Data Table 1 Doping Profile Information of Chip.}}
    \begin{tabular}{ccccc}\toprule
           Dopant&Region&$N_s$ (cm$^{-3}$)&   $y_0$ (µm)&$L_0$ (µm)\\ \midrule
           Phosphorus&N emitter&$1.58\times10^{20}$&   $13.10$&$21.71$\\ 
           Boron&P base&$9.55\times10^{17}$&   $0.00$&$104.90$\\  
 Aluminum& P base (beneath gate)& $2.39\times10^{15}$&  $39.93$&$166.81$\\ 
 Aluminum& P base (beneath cathode)& $6.17\times10^{15}$& $61.33$&$89.61$\\ 
 Phosphorus& N base& $3.50\times10^{16}$&  $0.00$&$45.93$\\ 
           Boron&P emitter&$6.92\times10^{17}$&   $0.08$&$2.56$\\\bottomrule
\end{tabular}
    \label{Table1}
\end{table}
\clearpage

\begin{figure}

    \centering
    \includegraphics[width=1\linewidth]{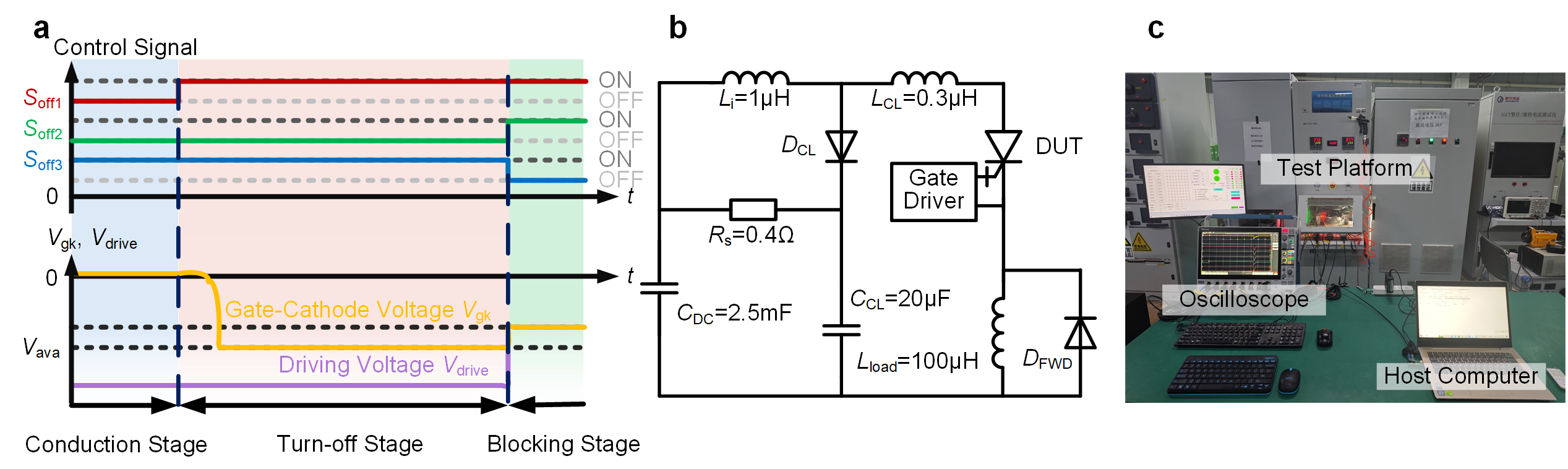}
    \caption{\textbf{Extended Data Fig.1  Paradigm Implementation and Experiment Platform Setup.}}
    \textbf{a, } Control algorithm of gate driver.
    \textbf{b,} Topology and circuit parameters of phase-leg circuit for switch-off operation safety verification. 
    \textbf{c,} Layouts of experiment platform.  
    \label{Extended data Fig1}
\end{figure}

\clearpage

\end{document}